\documentstyle[prl,aps,epsf,multicol]{revtex}

\begin{document}

\title{Comment on: ``High-Pressure Elasticity of $\alpha$-Quartz:
       Instability and Ferroelastic Transition''
       }

% \author{Martin H. M\"user and Philipp Sch\"offel}
\author{}
%
% \address{Institut f\"ur Physik, WA 331; Johannes Guntenberg-Universit\"at\\
%          55099 Mainz; Germany}
%
% \date{\today}
\maketitle

\begin{multicols}{2}

Gregoryanz et al.~\cite{gregoryanz00} have analyzed the stability of
$\alpha$-quartz under pressure in terms of the 
Born stability criterion (BSC)~\cite{born56}. The BSC requires
a positive definite elastic matrix constant $C_{ij}$
for stable elastic solids. 
It is strictly valid only at zero external pressure.
Important differences were discussed with theoretical predictions
by Binggeli and Chelikowsky~\cite{binggeli92}: The pressure regime
in which $\alpha$-quartz could be expected to be mechanically stable
if it was possible to suppress the quartz I - quartz II transition 
at $p = 22$~GPa 
was found to be significantly larger than predicted theoretically.
This conclusion is incorrect, because Gregoryanz et al. use an improper
generalization of BSC to non-zero pressures. The correct criterion
is to require that the matrix of Birch or 
stiffness coefficients~\cite{wallace72} $B_{ij}$
is positive definite rather than $C_{ij}$~\cite{wang93}.
Binggeli and Chelikowsky also state their BSC in terms of $C_{ij}$,
but as it seems to us, they accidentally calculated $B_{ij}$, e.g.,
they might have 
inverted the matrix
$ V \langle \delta \epsilon_i \delta \epsilon_j \rangle /k_B T$ with
 $\delta \epsilon_i$ being the thermal fluctuation of the strain tensor
around the equilibrium state at a given externally applied stress.
This inversion gives the Birch coefficients rather than $C_{ij}$; 
the equality $B_{ij} = C_{ij}$ only holds at zero external stress.
As we will show now by unpublished molecular dynamics simulations of
the present authors, the predictions of Binggeli and Chelikowsky 
are correct. We use the potential energy surface from 
Beest et al.~\cite{beest90}, which is known to be particularly well
suited to describe (elastic) properties of $\alpha$-quartz.
We focus on the BSC
\begin{equation}
B_3(C_{ij}) = (C_{11} - C_{12}) C_{44} -2 C_{14}^2 > 0,
\label{eq:b3}
\end{equation}
which is shown in Fig.~1.

In the inset of Fig.~1 it can be seen that the present
simulations reproduce the experimentally measured pressure dependence
of $B_3(C_{ij})$ (the central Fig.~3c of Ref.~[1])
fairly well.
We also want to note that we obtain the quartz I quartz II 
transition at the same pressure $p \pm 0.5$~GPa as in experiment.
These agreements justify the use of the BKS potential for 
$\alpha$-quartz under pressure. If $B_{\alpha\beta}$ replace $C_{\alpha\beta}$
in Eq.~\ref{eq:b3}, the behavior is similar to the one reported
by Binggeli and Chelikowsky ~\cite{binggeli92},
e.g., a nearly vanishing slope of $B_3(B_{ij})$ at small pressures is found,
and instability of $\alpha$ quartz can be expected to occur at 
$p \approx 25$~GPa. This is the correct generalization of 
BSC, which is equivalent to require that all eigenvalues of
$B_{ij}$ are positive. The minimum eigenvalue of $B_{ij}$ as obtained
in our simulations is shown in the main part of Fig.~1.

\begin{figure}[hbtp]
\begin{center}
\leavevmode
% \hbox{ \epsfxsize=110mm \epsfbox{min_eig.eps} }
\hbox{ \epsfxsize=60mm \epsfbox{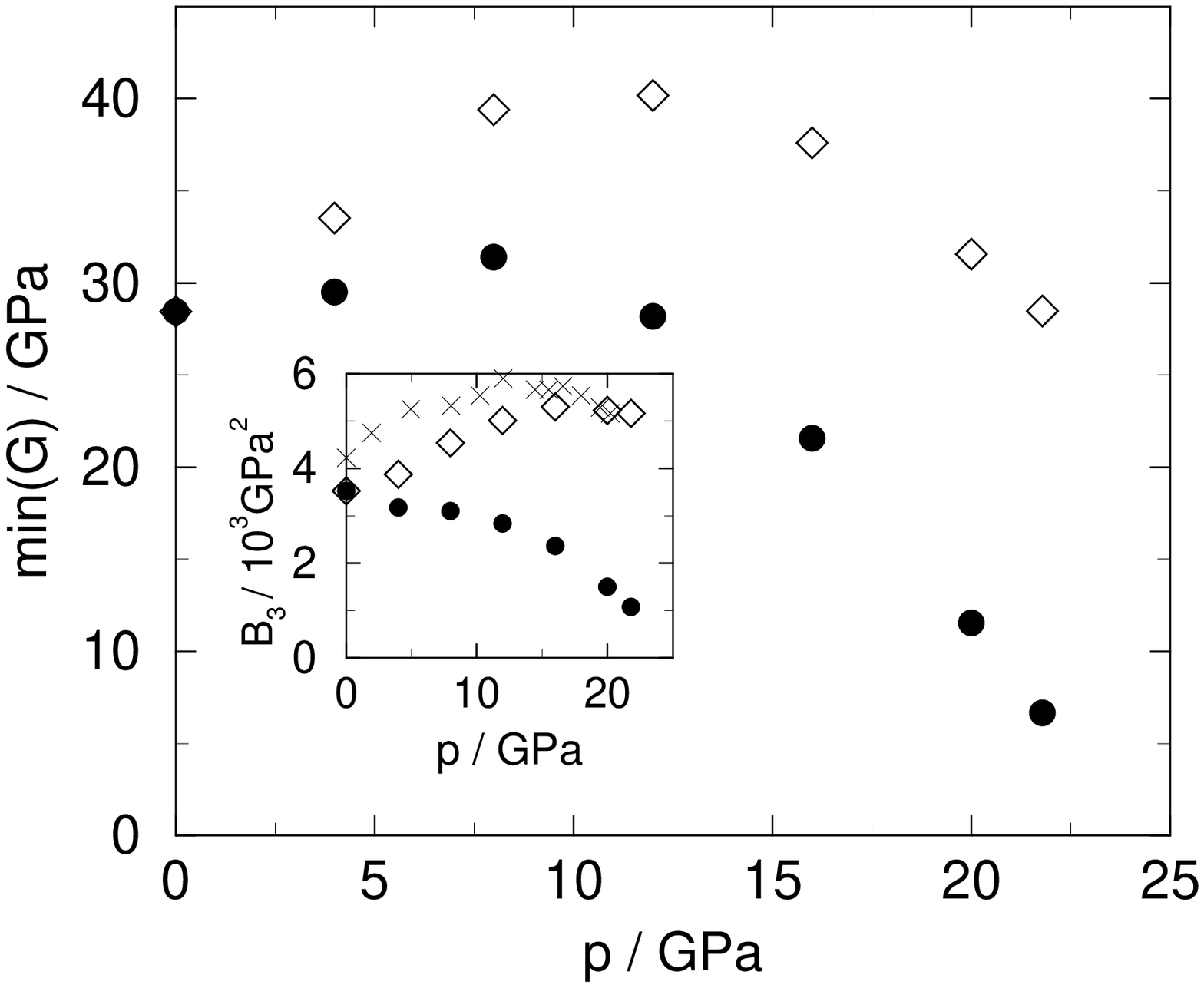} }
\begin{minipage}{8.5cm}
\caption{
Minimum eigenvalue of the matrix of the elastic constant (open diamonds)
and of the Birch matrix (filled circles) as a function of pressure.
Inset: Born stability criterion applied to elastic constants (diamonds)
and Birch coefficients (closed circles). Error bars are about symbol size.
Crosses represent experimental data from Ref.~[1].
\label{fig:stability}
}
\end{minipage}
\end{center}
\end{figure}

We further want to point out that the observation
of $\partial C_{44} / \partial p < 0$, which is discussed
explicitly in Ref.~\cite{gregoryanz00}, has no direct implication on
the mechanical stability. The only relevant influence of $C_{44}$
on mechanical stability in the case of an isotropic pressure $p$
enters via the properly generalized Born stability criterion
$B_3(B_{ij})$, see Eq.~(\ref{eq:b3}),
with $B_{44} = C_{44} - p$.\\

% Last, we wish to comment on the statement that the instability transition
% is ferroelastic. The weakening of an elastic constant elements with a
% changing 
% In our opinion, there should be at least one
% ferroelastic material involved in such a transition. While
% quartz I is definitely not ferroelastic, it is not known  to us
% whether or not quartz II is ferroelastic. Gregoryanz et al. do
% not provide an argument why this should be the case.

\noindent
We would like to thank K. Binder for useful discussions.\\

\noindent
M. H. M\"user and P. Sch\"offel\\
\hspace*{5mm} Inst. f\"ur Physik, WA 331 \\
\hspace*{5mm} Johannes Gutenberg-Universit\"at\\
\hspace*{5mm} 55099 Mainz, Germany

\end{multicols}


\begin{thebibliography}{99}

\bibitem{gregoryanz00}
E. Gregoryanz, R. J. Hemley, H. K. Mao, and P. Gillet,
Phys. Rev. Lett. {\bf 84}, 3117 (2000).

\bibitem{born56}
M. Born and K. Huang, {\it Theory of Crystal Lattices}
(Clarendon, Oxford, 1956).

\bibitem{binggeli92}
N. Binggeli and J. R. Chelikowsky,
Phys. Rev. Lett. {\bf 69}, 2220 (1992).

\bibitem{wallace72}
D. C. Wallace, {\it Thermodynamics of Crystals}
(Wiley, New York, 1972).

\bibitem{wang93}
J. Wang, S. Yip, S. R. Phillpot, and D. Wolf,
Phys. Rev. Lett. {\bf 71}, 4182 (1993).  

\bibitem{beest90}
B. van Beest, G. Kramer, and R. van Santen,
Phys. Rev. Lett. {\bf 64}, 1955 (1990). 

\end{thebibliography}
\end{document}